\documentclass[aps,prl,twocolumn,groupedaddress,10pt,showpacs]{revtex4-1}
\usepackage{graphicx}
\usepackage{amsmath,amssymb,bm}
\usepackage{url}

\begin{document}

\title{Econophysics on Real Economy\\[5pt]
-- The First Decade of the Kyoto Econophysics Group --}

\author{Hideaki Aoyama}
\email[]{hideaki.aoyama@scphys.kyoto-u.ac.jp}
\affiliation{Department of Physics, Kyoto University, Kyoto 606-8501, Japan}
\author{Yoshi Fujiwara}
\email[]{yoshi.fujiwara@gmail.com}
\affiliation{ATR Laboratories, Kyoto 619-0288, Japan}
\author{Yuichi Ikeda}
\email[]{Yuichi.IKEDA@iea.org}
\affiliation{Hitachi Ltd, Hitachi Research Laboratory, Ibaraki 319-1221, Japan\\
{\rm and} International Energy Agency, 
75739 Paris Cedex 15, France}
\author{Hiroshi Iyetomi}
\email[]{hiyetomi@sc.niigata-u.ac.jp}
\affiliation{Department of Physics, Niigata University, Niigata 950-2181, Japan}
\author{Wataru Souma}
\email[]{souma.wataru@nihon-u.ac.jp}
\affiliation{College of Science and Technology, Nihon University, Chiba 274-8501, Japan\\
{\rm and} Institute of Economic Research,  Hitotsubashi University, Tokyo 186-8603, Japan}

\date{June 29th, 2010: KUNS-2275}

\begin{abstract}
Research activities of Kyoto Econophysics Group is reviewed.
Strong emphasis has been placed on real economy.
While the initial stage of research was a first high-definition data analysis 
on personal income, it soon progressed to firm dynamics,
growth rate distribution and establishment of Pareto's law 
and Gibrat's law. It then led to analysis and simulation of firm dynamics on economic 
network. Currently it covers a wide range of dynamics of
firms and financial institutions on complex network,
using Japanese large-scale network data, some of which are not
available in other countries. Activities of this group for publicising
and promoting understanding of econophysics is also reviewed.
\end{abstract}

\maketitle

\section{Introduction}
Physics community in Japan has been blessed with open-mindedness
about application of physics methods and attitudes to other areas 
of research.
 
One of the earliest expression of such an ideal was  by
Torahiko Terada, a leading physicist and a well-respected writer of that time, 
who wrote the following defence
of bringing physics into  biology in 1933 \cite{torahikoterada}:

\begin{quote}
When making a statistical analysis of a large number of human individuals we may properly regard it as a mere conglomeration of inorganic material, and altogether neglect individual free will.
Indeed, it is now clear that pure physical problems, such as 
the density of particles in a colloidal matter, may with propriety be compared to statistics of a purely physical nature, such as the ``density" or ``average speed" of persons walking along street.
\dots \
It is sheer folly to dismiss such insights as heresy simply because they are incompatible with the dogma that ``living creatures cannot be understood by Physics". Such absurdities remind us that no ignorant amateur poses so serious a threat to progress as a scientist unaware of the nature and goal of their discipline.
\end{quote}

Hideki Yukawa, who won a Nobel Prize in Physics for his meson theory in 1949 
for the first time in Japan,
encouraged his disciples to spread the physics research to
various areas, including biology and cosmology, which all 
blossomed in the following years. One of his favorite quote from Kyogen
(Japanese traditional comedies, in the style of Noh)
is the following \cite{waranbesou}:

\begin{quote}
A patient mind looks everywhere for signs of thoughts and things of worth; 
The shallowest stream runs sparkling over secrets far beneath the earth.
\end{quote}

Econophysics was introduced to Japanese physics community by Hideki Takayasu (Sony CSL),
whose pioneering work may be, 
the very first mathematical modeling and simulation of stock market in 1992
\cite{takayasu1992statistical},
and application of the Langevin equation to the stochastic process and the
derivation of the power law \cite{takayasu1997stable}, among others.
He also publicised the approach and fruits of econophysics widely in physics community
through lectures and organization of conferences.
The present authors, who studied various areas of physics;
elementary particle physics (H.A. and W.S.), 
cosmology (Y.F.), 
condensed matter physics (H.I.)
and high-energy and nuclear physics (Y.I.),
met econophysics under his influence.
Since then, Kyoto Econophysics Group has been conducting research
with emphasis on real economy, and has published two books in Japanese \cite{pfbook,ecbook},
and a book in English is forthcoming in August 2010 \cite{bible2010}.
In the following, we shall review our research activities of the last decade
and will conclude it with the future prospect.

\section{Pareto's law and Gibrat's law}
The beginning of the econophysics research at Kyoto
may be the empirical study of the personal income in 2000 \cite{aoyama2000}, 
which is the first high-definition study, covering some eighty thousand top 
income-earners in Japan.
It came about as a result of a chance finding of a data CD that contained the list
of the income, address, phone numbers, etc.\ of those people
by one of the authors (H.A.) at a PC shop in Akihabara, Tokyo.
(Such information were posted at the tax offices scattered all over in Japan
and were collected by a database company at that time, but is no longer available
for the protection of personal information.)
This resulted in clear identification of the
Pareto's (power) law, followed by several related works
\cite{souma2001universal,souma2002physics},
earned its position as a logo of a conference \cite{kolkata1}
\footnote{{http://www.springerlink.com/content/n21751/cover-large.gif}}

Y.\ Fujiwara joined this group in the midst of this line of work,
after working on stock markets:
some of which were on self-similarity and multifractality in stock price fluctuations
\cite{fujiwara2001cgs} and a spin model of stock market \cite{kaizoji2002dpt}.

The research thus started soon lead to study of the growth rate, 
the ratio of the income of a year and the income of the next year for each person,
which was a collaborative work with economists, Masanao Aoki (UCLA) and 
Taisei Kaizoji (ICU) \cite{fujiwara2003growth},
as we recognized importance of collaboration between physicists and economists.
It may be noteworthy that H.A.\ knew that when physicists enter areas outside physics
such a collaboration with experts is almost a must for fruitful results,
from his earlier work with John Constable, a Cambridge-educated English critic/poet,
on linguistics \cite{aoyama1999word,constable1999word,constable2001testing}. 

Soon, it flourished to various research covering firms as well:
establishment of the Gibrat's law in high-definition data 
\cite{aoyama2003growth,aoyama2004kinematics},
study of European firms, with Italian economists, Corrado di Guilmi and Mauro 
Gallegati \cite{fujiwara2004pareto,fujiwara2004gibrat},
further study of the firm-size distribution \cite{kaizoji2006res},
study of small-to-medium firms using the Credit Risk Database (CRD) in 
Japan \cite{fujiwara2006gfs},
study of firm bankruptcy, the life-time distribution and a
simulation model (Delli~Gatti-Gallegati-Palestrini) \cite{fujiwara2004zlf},
and two-factor stochastic model, with economist Nirei,
to explain personal income distribution in Japan and U.S.
\cite{nirei2004income,souma2005empirical,nirei2007two}.

In the meantime, 
Y.\ Ikeda had been studying applying the real-option theories in financial engineering and game theoretical approach for risk assessment in multiple-business environment
\cite{Ikeda200487,Ikeda2006:210,IKEDAYuichi:2006-153},
which motivated him to join collaborative work with the other authors. 
He brought research funding from Hitachi to us, which was a key ingredient
for the formation of the Kyoto group.
He also continued his own line of research, which was connected to his company's interest,
including a study of project risk using Bloomberg data \cite{KAWAMOTOShigeru:2006-09-01},
and a work on correlation between R\&D investment and sales-growth \cite{Tomita2008}.

\section{Economic Networks}

{\it Interactions\/} among heterogeneous agents are crucial for
understanding of emergent phenomena at macroscopic levels of real economy.
The econophysics community has recently witnessed considerable development
of complex networks for quantifying structures, temporal changes
of various economic networks, their relevance in macroeconomics,
and so forth. See the Econophysics-Kolkata III in 2007 \cite{kolkata3},
a collection of monographs \cite{chakrabarti2006}, and also a recent
review \cite{schweitzer2009en} with references therein.

Our early papers in \cite{souma2001swe,souma2003wealth} studied the
small-world effects of a Bouchaud-M\'ezard model for the stochastic
dynamics of wealth distributed among agents on a network. 
We then examined real data of economic
networks in Japan to investigate complex networks and economics
\cite{souma2003complex}, structure and change of shareholding
relationships between firms
\cite{souma2004random,souma2005shareholding,souma2006change,ss},
growth of firms and networks \cite{fujiwara2006gfn}, and
a study on correlation between firm's financial states and its
characteristics in a transaction (supplier-customer) network
\cite{souma2006cbn}.

We also developed agent-based models to understand heterogeneous
interactions among economic agents, particularly on a transaction
network, including response of firm agent network to exogenous shock
\cite{ikeda2007rfa}, simulation for chain of bankruptcy
\cite{ikeda2007asc}, firm dynamics with parameters estimated by
financial and transaction data analysis \cite{ikeda2007qab}, and a
study on correlation between firms' performance and characteristic
properties in the network \cite{ikeda}.
Dynamics of coupled balance-sheets of firms
({\it \`a la} Delli~Gatti-Gallegati-Palestrini)
was also studied in \cite{Iyetomi2005,Iyetomi2009}, and the
paper \cite{iino2009mor} focused on the model of relation between
transaction network and production activity of firms.

Recently, in addition to commercially available data, we had a quite
unique opportunity to collaborate with a leading credit research
agency in Japan which regularly gathers credit information on most of
active firms. Through collaboration, we examined a nation-wide
transaction network comprising a million firms and millions of
supplier-customer links. The studies on network structure, uncovering
community (cohesive groups), visualization and graph drawing are
included in \cite{fujiwara2008lss,fujiwara2009vls} and more
intensively in
\cite{iyetomi2009elucidation,kamehama2010vis,iino2010comm}. This
direction would give an insight into industrial structure of the
economic system of the nation as well as instability {\it via\/}
chain of failures \cite{fujiwara2008cfb}.

Financial networks are no less important than transaction
networks. Focusing on banks and firms relationships, we also extended
our study on financial networks in \cite{masi2008ajc,fujiwara2009stc}.

In relation to productivity and production function discussed in the
following section, we can mention the relation between networks of
firms and the theory of ``ridge'' in the space of production function
\cite{souma:nfa}.

\section{Labour Productivity}
Economists Hiroshi Yoshikawa (University of Tokyo) and Masanao Aoki
proposed a model of labour productivity distribution based on
statistical physics ideas in 2003 \cite{aokiyoshikawa,yoshikawa2003}.
This spurred interests among us and we started analysing Japanese
data in 2007 in collaboration with the two economists.
It soon lead to finding of the power-law distribution
and the superstatistics theory of labour distribution.
The relevant paper appeared in 2008
as arXiv:0805.2792v1 [q-fin.GN] and RIETI Discussion Paper 08-E-035,
but published only in 2010 \cite{aoyama2010disp} after a long delay 
in reviewing process and rejection of the traditional mainstream economists.
In any case, this line of work soon lead to further mathematical disposition \cite{aoyama2008b},
study of the Period 1996-2006 \cite{souma2008x},
international comparisons \cite{ikeda2009iclpd},
study of Japanese manufacturing and non-manufacturing sectors \cite{economics2009-22},
and use of large-scale data of firm's financial statements \cite{ikeda-analysis}.

As a by-product of the study of productivity, we proposed a new concept of 
{\it Production Copula},
which is a multi-variate statistical distribution that should
replace the traditional `production function' \cite{iyetomi2009pc}.

Another line of work was created from the above:
detailed study of productivity distribution, using the huge CRD data lead to
a new finding about the medium range of the productivity, not explained by
the superstatistics \cite{yoshi2010stochastic}.
Furthermore, new scaling laws were discovered and new mathematical theory was developed,
which opens up a whole new path to the formulation of statistical physics of economy 
\cite{aoyama2010micromacro},

\section{Business Cycles}

Business cycles are a long-standing basic issue for macroeconomics. Even whether a business cycle exists or not is controversial among economists.

Very recently we have analyzed~\cite{iyetomi2009a} business cycles in Japan using the indices of industrial production (IIP), an economic indicator which measures the current conditions of production activities over the nation on a monthly basis. Careful elimination of statistical noises based on the random matrix theory (RMT) enabled us to extract two dominant factors each of which possess its own economic characteristics, aggregate demand and inventory adjustment, respectively. The two factors collaborate in giving rise to business cycles with periods of 40 and 60 months which are hidden behind complicated stochastic behaviors of the indices. Compiling all of the results in~\cite{iyetomi2009a} suggests that the major cause of business cycles is real demand shocks. In passing we note that Japanese business cycles were studied also from the point of view of a chaotic theory through reconstruction of attractor for the GDP time series~\cite{Nishigaki:2007}. 

The genuine correlation matrix for the IIP thus obtained elucidates interindustry correlations in a statistically meaningful way~\cite{iyetomi2009b} as well as the business cycles. The fluctuation-dissipation theory was invoked to elucidate input-output industrial correlations quantitatively. And we observed distinctive external stimuli on the Japanese economy exerted by the recent global economic crisis. Those stimuli were derived from residual of moving-averaged fluctuations of the IIP left over after subtracting the long-period components due to the business cycles. 

The RMT is also useful for construction of stock portfolio
\cite{fujiwara2006application}. The idea of the RMT was then extended~\cite{nakayamai2009rmt} to detect genuine dynamical correlations between price fluctuations of different stocks. A data set of one-day returns of 557 Japanese major stocks for the 11-year period from 1996 to 2006 was used for this study. Comparison of the eigenvalues of the empirical dynamical correlation matrix with the corresponding results for random data demonstrates existence of collective motions of the stock prices with periods well over days.

\section{Extra-research Activities and Future Prospect}
In addition to the research activities reviewed above, we have been working
on outreaching to physicists and economists,
publicizing the merits and viewpoints of econophysics, 
and at the same time encouraging discussions and collaborations between econophysicists
in Japan as well as abroad.

We have ran four domestic econophysics conference at
Yukawa Institute for Theoretical Physics at Kyoto University, in 2003, 2005, 2007 and 2009.
Attendance was about 70 to 90, with undergraduate and graduate students at steady increase.
The third conference report was published as a Supplement of {\it Progress of
Theoretical Physics}, 
while the Proceedings of the other conferences
were published in {\it Elementary Particle Physics Reports} and
{\it Condensed Matter Physics Reports}, both of which are publications of
the Yukawa Institute in Japanese, circulated widely to the Japanese theoretical 
physics community.

Further we are currently running a series of Japanese articles 
{\it Introduction to Econophysics} 
in {\it the Kinyu-Zaisei Business Journal} (Jiji Press, Ltd. Tokyo),
where in addition to our own articles we have been inviting articles from every aspect 
of econophysics research in Japan. Since this weekly journal is 
widely read by high-ranking government officials and presidents 
of financial institutions,
we expect that understanding of ``econophysics'' is deepened for them through this effort.

Graduate students studied under us for Master's degree in physics are not many. 
But they have obtained positions at the cabinet office of the government of Japan, 
several financial institutions and other firms, which imply that 
econophysics background is welcomed.
University positions held by the present authors are those of
physics and natural-science departments and not of econophysics, which
makes it difficult to accept graduate students who wish to study econophysics.
But econophysics is attracting more and more interests of students, as demonstrated 
by their increasing attendance in our conferences.  
We thus trust that this situation could be remedied in future.

All in all, we conclude that the prospect for the econophysics is very high in Japan:
Econophysics is expected to make huge impact on economics, with its new ideas and 
approaches suitable for economic phenomena, imported from various areas of physics.

\begin{acknowledgments}
\section{Acknowledgements}
The authors are grateful to Bikas K. Chakrabarti and Anirban Chakraborti
for inviting us to write this review.
Many people have assisted us in research reviewed in this paper
and we thank them all.
Some of them are;
Masanao Aoki (Los Angeles),
John Constable (London),
the late Hirokazu Fujisaka,
Mauro Gallegati (Ancona),
Corrado Di Guilmi (Ancona),
Shigeru Hikuma (Tokyo),
Yasuyuki Kuratsu (Tokyo),
Hiroyasu Inoue (Osaka),
Taisei Kaizoji (Tokyo),
Makoto Nirei (Tokyo),
Makoto Nukaga (Tokyo),
Hideki Takayasu (Tokyo),
Misako Takayasu (Tokyo),
Schumpeter Tamada (Hyogo),
Hiwon Yoon (Tokyo),
and
Hiroshi Yoshikawa (Tokyo).
The present work is supported in part by 
{\it the Program for Promoting Methodological Innovation in Humanities and Social Sciences by Cross-Disciplinary Fusing} of the Japan Society for the Promotion of Science, 
{\it Grant-in-Aid for Scientific Research (B)} 20330060 (2008-10) and 22300080 (2010-12),
{\it Invitation Fellowship Program for Research in Japan (Short term)} ID No.S-09132 of the Ministry of Education, Science, Sports and Culture, Japan,
{\it Hitachi, Ltd.}, 
{\it Hitachi Research Institute}, 
and 
{\it The Research Institute of Economy, Trade and Industry (RIETI)}.
\end{acknowledgments}


%

\end{document}